\RequirePackage{fix-cm}

\documentclass[smallextended]{svjour3}       
\smartqed  

\usepackage{graphicx}
\usepackage{dcolumn}
\usepackage{bm}
\usepackage{textcomp}
\usepackage{verbatim} 
\usepackage{datetime}
\usepackage{color}

\usepackage{color}
\definecolor{darkblue}{rgb}{0,0,0.6}
\usepackage[linkcolor=darkblue, citecolor=darkblue,colorlinks=true, breaklinks=true]{hyperref}

\begin{document}

\title{{\color{blue}Time-domain Brillouin Scattering as a Local Temperature Probe in Liquids}}

\author{Ievgeniia Chaban \and
Hyun D. Shin \and
Christoph Klieber \and
R\'emi Busselez \and
Vitalyi E. Gusev \and
Keith A. Nelson \and
Thomas Pezeril
}

\institute{I. Chaban \at
Institut des Mol\'ecules et Mat\'eriaux du Mans, UMR CNRS 6283, Le Mans Universit\'e, 72085 Le Mans, France \\
\email{ievgeniia.chaban@univ-lemans.fr}
\and
H. D. Shin \at
Department of Chemistry, Massachusetts Institute of Technology, Cambridge, MA 02139, USA
\and
C. Klieber \at
Department of Chemistry, Massachusetts Institute of Technology, Cambridge, MA 02139, USA \\
\emph{Present address: }{ EP Schlumberger, 1 rue Henri Becquerel, 92140 Clamart, France}
\and
R. Busselez \at
Institut des Mol\'ecules et Mat\'eriaux du Mans, UMR CNRS 6283, Le Mans Universit\'e, 72085 Le Mans, France
\and
V. Gusev \at
Laboratoire d'Acoustique de l'Universit\'e du Maine, UMR CNRS 6613, Le Mans Universit\'e, 72085 Le Mans, France
\and
Keith A. Nelson \at
\email{kanelson@mit.edu}\\
Department of Chemistry, Massachusetts Institute of Technology, Cambridge, MA 02139, USA
\and
T. Pezeril \at
\email{thomas.pezeril@univ-lemans.fr}\\
Institut des Mol\'ecules et Mat\'eriaux du Mans, UMR CNRS 6283, Le MansUniversit\'e, 72085 Le Mans, France}

\date{Received: date / Accepted: date}

\maketitle

\begin{abstract}
{{\color{black}We present results of time-domain Brillouin scattering (TDBS) to determine the local temperature of liquids in contact to an optical transducer. TDBS is based on an ultrafast pump-probe technique to determine the light scattering frequency shift caused by the propagation of coherent acoustic waves in a sample. Since the temperature influences the Brillouin scattering frequency shift, the TDBS signal probes the local temperature of the liquid. Results for the extracted Brillouin scattering frequencies recorded at different liquid temperatures and at different laser powers - i.e. different steady state background temperatures- are shown to demonstrate the usefulness of TDBS as a temperature probe. This TDBS experimental scheme is a first step towards the investigation of ultrathin liquids measured by GHz ultrasonic probing.}}
\PACS{PACS code1 \and PACS code2 \and more}
\subclass{MSC code1 \and MSC code2 \and more}
\end{abstract}


\section{Introduction}

{{\color{black}Femtosecond laser pump-probe techniques are often used to investigate thermal transport in materials such as metals, semi-conductors, insulators, gases, liquids and at solid-liquid-gas interfaces. These techniques include measuring the transient sample temperature rise following the partial absorption of a femtosecond laser pump pulse through monitoring the change in optical reflectivity by an optically time-delayed probe. Analysis of the recorded waveforms is typically done through comparison to models in order to retrieve the thermal characteristics of the sample, such as thermal conduction, diffusivity or even Kapitza interfacial thermal resistance \cite{Cahill2004,Schmidt2008,Schmidt2008b,Cahill2014,Schmidt2009}. Time-domain Brillouin scattering (TDBS) is also based on ultrafast lasers for optical ultrasound excitation and detection in the materials under study. It is a well suited technique for the measurement of elastic, visco-elastic and optical properties of ultrathin transparent solid or liquid samples at ultrasound GHz to THz frequencies \cite{Lin1991}. TDBS has been applied for the examination of diverse phenomena such as non-linear acoustic waves in solids and liquids \cite{Klieber2015,Bojahr2012}, spatial mechanical inhomogeneities in solids \cite{Mechri2009,Nikitin2015}, and even GHz transverse acoustic phonons in viscoelastic liquids \cite{pezeril09,pezeril12,KHP+13,pezeril16}.

We present a tabletop pump-probe method which enables the measurement through TDBS of the local temperature in liquids in contact to an optical transducer. We  demonstrate the performances of TDBS as a contactless local temperature probe with liquid glycerol, a well-known and well-characterized prototypical glass-forming liquid. We also study octamethylcyclotetrasiloxane (OMCTS), a confined liquid prototype, with the interest for future experiments related to molecular confinement in ultrathin liquid layers where the understanding of thermal effects remains elusive.}

\section{Experimental technique}

For sample construction, the liquid under study was squeezed in between two flat optical quality substrates with one of them having a metallic thin film deposited on it which was in contact with the liquid, as sketched in Fig. \ref{fig:sample_holder}. The metallic film, which, depending on the experimental configuration, is either a 40~nm chromium film or a 80~nm aluminium film, was acting as a photoacoustic transducer and the liquid thickness was more than 10~microns. As liquids we used glycerol (Acros Organics$^\textrm{\textregistered}$, 99+\% purity) and OMCTS (Fluka$^\textrm{\textregistered}$, 99+\% purity) which were prepared by forcing them through several linked 0.2~$\mu$m teflon millipore filters to remove dust particles. {\color{black}The fully assembled liquid sample cell was installed inside a cold finger cryostat which was temperature controlled through a feedback sensor directly attached to its heat sink (labeled Sensor A in Fig. \ref{fig:sample_holder}). The sample temperature was measured by an additional Peltier temperature sensor (labeled Sensor B in Fig. \ref{fig:sample_holder}) which was attached to the sample holder mount}, about 1~to~2~cm away from the experimental volume, {\color{black}the closest place for a convenient installation. At each temperature, the sample was given sufficient time to equilibrate before data acquisition. This time was estimated upon equilibration of the cryostat temperature sensors A and B and from repeated TDBS measurements until convergence to a reproducible steady state TDBS signal.}

\begin{figure}[t!]
\centering
\includegraphics[width=9cm]{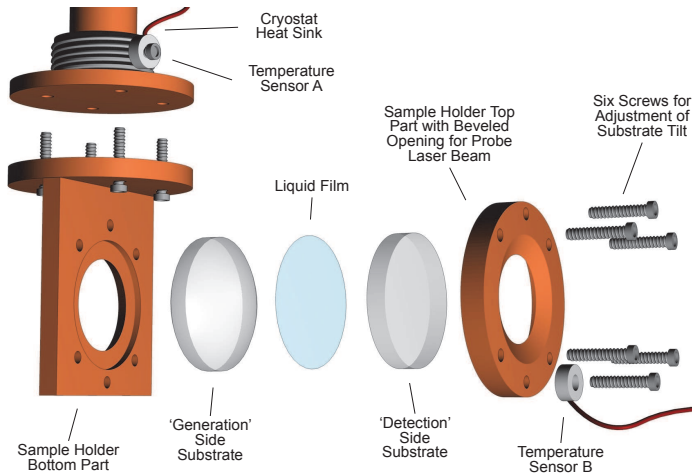}
\caption{Illustration of the sample holder used to construct liquid samples of different thicknesses. A copper jig was used to hold two optically transparent substrates with the sample liquid sandwiched in between. The whole construction is attached to a heat sink of a cryostat with one temperature sensor (A) at the heat sink and a second (B) attached to the jig to monitor the sample temperature as closely to the sample as possible.}
\label{fig:sample_holder}
\end{figure}

\begin{figure}[t!]
\centering
\includegraphics[width=9cm]{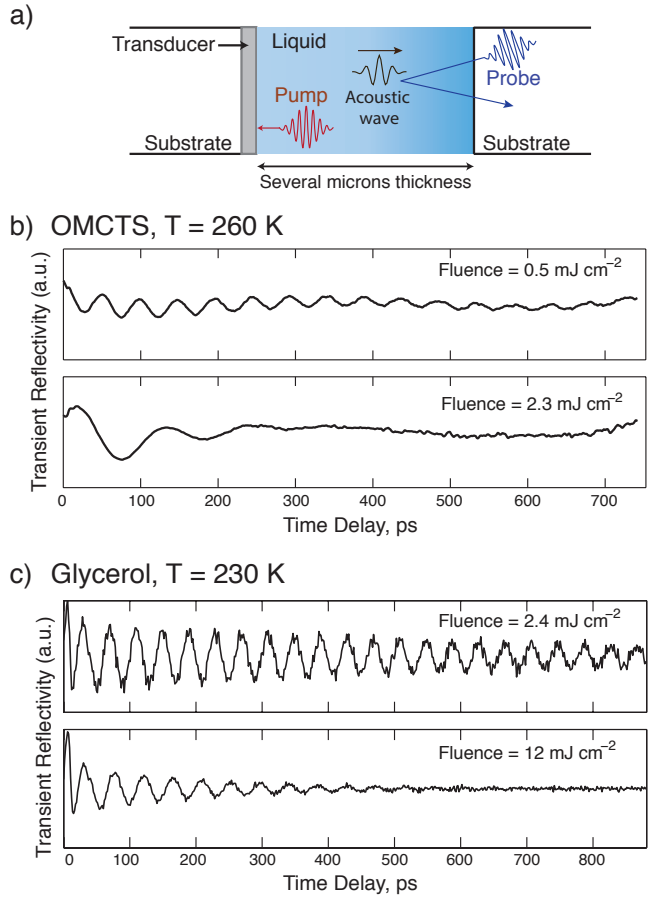}
\caption[Fig1]{(a) Sketch of sample with a liquid squeezed in between two flat, optically grade substrates. One substrate holds a metallic thin film serving as photoacoustic transducer. Laser irradiation launches the acoustic waves through the metallic transducer film which are transmitted through the transparent liquid where they are detected through TDBS by means of a time-delayed laser probe pulse. For several different pump fluences and at different temperatures of the cryostat we recorded transient reflectivity data for OMCTS (b) and glycerol (c). For a fivefold change in laser fluence, the Brillouin scattering frequency drastically changes in case of OMCTS which indicates melting through local laser thermal heating. For a similar change in pump fluence for the glycerol sample, the Brillouin frequency changes by over 10\%.}
\label{fig1}
\end{figure}

The optical experiment uses the TDBS technique, suitable for the study of the temperature and frequency dependences of ultrafast mechanical dynamics in liquids in the lower GHz frequency range \cite{pezeril09,pezeril12,KHP+13,pezeril16,Kli10,Shelton2005,Maznev11}. Measurements were carried out using an ultrafast optical pump-probe experimental setup as illustrated in Fig. \ref{fig1}(a). The laser pulses were generated by a femtosecond Ti-Sapphire Coherent RegA 9000 regenerative amplifier operating at a repetition rate of 250~kHz outputting 160~fs pulses with a central wavelength of 790~nm. The laser pulses were separated into two beams with a 790~nm pump beam synchronously modulated at 50~kHz frequency by an acousto-optic modulator (AOM) and focused on the surface of a metallic photo-acoustic transducer film with a gaussian spatial beam profile of FWHM $\sim$100~$\mu$m. {\color{black}For an enhanced signal to noise ratio, the pump modulation frequency has been chosen as a submultiple of the laser repetition rate in order to obtain a stable number of pump pulses that are chopped on each cycle.} The second beam, used as the probe, was a much less energetic beam which was frequency doubled by a BBO crystal to 395 nm. It was time-delayed and tightly focused on the sample surface at normal incidence with a spot size smaller than 20 $\mu$m where it spatially overlapped with the pump spot. A photodiode, coupled to a lock-in amplifier synchronized to the 50~kHz pump modulation frequency, recorded the reflected probe beam and hence measured the transient differential reflectivity $\Delta$R(t) as a function of delay between pump and probe beams. Upon transient absorption of the 790~nm pump pulse over the optical skin depth of the metallic thin film, laser excited acoustic pulses were transmitted across the interface into the adjacent transparent liquid. The out-of-plane acoustic propagation of the strain pulses in the transparent liquid medium leads to the occurrence of TDBS oscillations in the transient reflectivity signal, see Fig. \ref{fig1}(b) and (c). As in any Brillouin scattering process, the frequency $\nu$ of these oscillations is related to the ultrasound velocity $v$ of the liquid, to the probe wavelength $\lambda$, to the refractive index $n$ of the medium, and to the back-scattering angle $\theta$ through
\begin{equation}\label{eq:Brillouin1}
\nu = 2 \ n \ v \cos \theta / \lambda.
\end{equation}
The acoustic velocity and the index of refraction of the liquid and, as a consequence, the TDBS itself are influenced by many external conditions, such as the ambient temperature and the laser fluence. Therefore TDBS is sensitive to a local temperature modification of the scattering liquid medium. In fact, the Brillouin scattering frequency detected in the time-domain reflectivity signal monitors any change of the temperature distribution, which appears as a modification of the detected Brillouin oscillation frequency. {\color{black}As in classical interferometric processes, the characteristic TDBS sensitivity length is given by $\lambda/2n$. This sensitivity length is, in most cases, in the range of 100-200 of nanometers.} It means that any change of the overall temperature or temperature distribution in a liquid volume as small as a couple of pico-liters (200~nm$\times$probe spot surface of 100~$\mu$m FWHM diameter) can be detected from TDBS.

The laser pump pulse can cause permanent damage or irreversible sample modification at a given fluence threshold, as in any optical pump-probe experiment. This effect can be experimentally observed once the recorded data become fluence dependent such as the excitation of shock waves at high laser fluences \cite{Klieber2015} which reveals the departure from the linear acoustic regime to the non-linear acoustic regime. It can be a consequence as well of a local temperature rise caused by cumulative heating of the sample from the multiple laser pump pulses which brings the sample into a steady state temperature regime correlated to the laser pump fluence \cite{Cahill2004,Schmidt2008,Schmidt2008b,Cahill2014,Schmidt2009}. Fig. \ref{fig1}{\color{black}(b) and (c)} displays recorded data obtained in OMCTS and glycerol at a temperature of 260~K and 230~K, respectively, as indicated by the Peltier temperature sensor, at different laser pump fluences. As seen in Fig. \ref{fig1}(b), a fivefold change in the laser pump fluence induces a drastic change in the Brillouin oscillations frequency, from 20.5~GHz to 8.8~GHz, and in the attenuation rate. The Brillouin frequency of 20.5~GHz matches the Brillouin frequency in the glassy state whereas the Brillouin frequency of 8.8~GHz matches the Brillouin frequency in the liquid state, see Fig. \ref{fig2}(a), indicating a major modification of the OMCTS sample which experiences melting mediated by the cumulative heating of the multiple laser pump pulses. Similarly, the Brillouin frequency in glycerol changes from 25.1~GHz to 22.8~GHz, for a fivefold modification of the laser pump fluence. Similarly, the attenuation rate evolves with a modification of the laser fluence, however, its pertinence is out of scope of the manuscript which focus mainly on the analysis of the Brillouin frequency versus laser fluence or temperature.

An important aspect to consider is that when the liquid is driven with large enough strain pulses to produce shock wave formation as in \cite{Klieber2015}, the Brillouin frequency will increase with an increase of the laser pump fluence, which is opposite to our current experimental observations. Therefore, we have neglected the effect of non-linear shock waves and solely assumed cumulative laser heating as the main mechanism responsible for the evolution of the Brillouin frequency in our present measurements. In the following, we will describe how to calibrate the measured Brillouin frequency in the studied liquid in order to employ TDBS as a specific temperature sensor.

\section{Results}

\begin{figure}[t!]
\centering
\includegraphics[width=9cm]{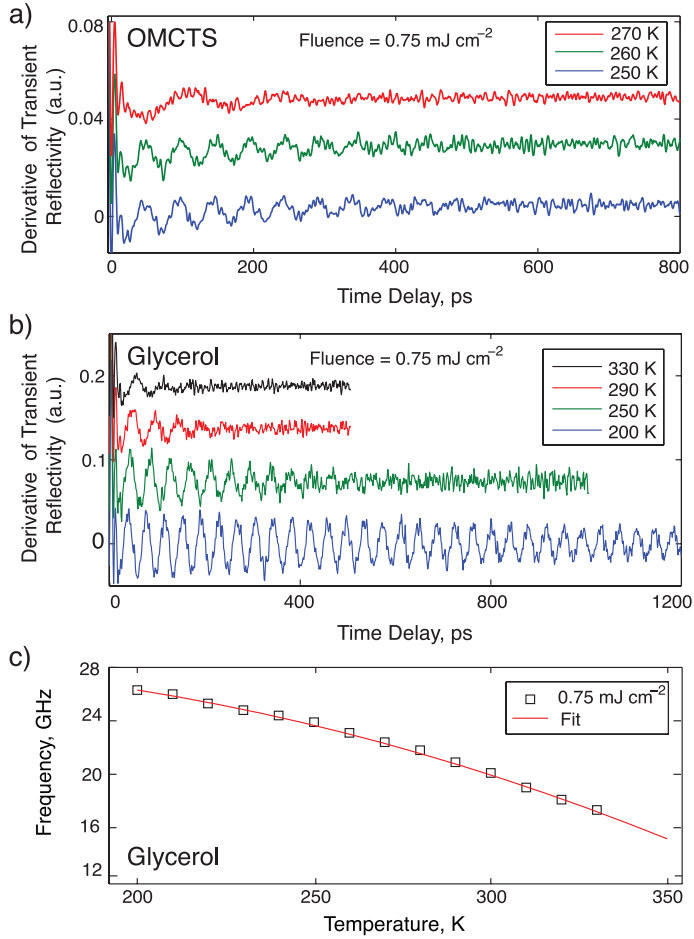}
\caption[Fig2]{Time derivative of recorded transient reflectivity signals obtained in (a) OMCTS and in (b) glycerol at different temperatures of the cryostat and at a given laser pump fluence of 0.75 mJ.cm$^{-2}$. At such relatively low fluence, the overheating caused by the multiple laser pump pulses is moderate. (c) Temperature dependent Brillouin frequency in glycerol. The temperature calibration curve displayed in (c) can be used to estimate the absolute liquid temperature from the measured Brillouin frequency. {\color{black}The experimental uncertainties in (c) are estimated to lie below 0.1~GHz}.}
\label{fig2}
\end{figure}

The temperature calibration measurement takes advantage of the fact that the Brillouin frequency in a liquid is strongly temperature dependent. Therefore, the Brillouin frequency can be used as a probe to determine the absolute temperature in the experimentally investigated local region of the liquid. Since the laser pump itself can affect the Brillouin frequency at high fluences, the measurement data shown in Fig.~\ref{fig2} were obtained at sufficiently low pump fluence such that the effect of cumulative heating is moderate. Figure~\ref{fig2} shows the derivative of the recorded transient reflectivity change recorded at several temperatures in OMCTS and glycerol, with a sample structure as sketched in Fig.~\ref{fig1}(a), composed of a silicon substrate holding a 40~nm chromium film in contact with the liquid and a transparent cover glass substrate. In both cases, for both liquids, the temperature influences the Brillouin frequency and the attenuation rate, as shown in Fig.~\ref{fig2}(a) and (b). {\color{black}OMCTS displays a sharp crystalline to liquid phase transition at 260~K and the Brillouin frequency variation across the transition is way more abrupt than for glycerol which is an intermediate fragile glass former with a smooth glass transition temperature around T$_g$~=~186~K. A sharp temperature transition is not appropriate for the TDBS temperature determination of a liquid sample over a wide temperature range. For this physical reason, our TDBS technique applied for the temperature determination of samples is more adapted for glass former liquids. In the following, we detail the meticulous TDBS temperature calibration in the case of glycerol.} For each temperature measurement in glycerol, such as the ones displayed in Fig.~\ref{fig2}(b), the frequency $\nu$ of the Brillouin oscillations observed in the transient reflectivity signal were fitted following a sinusoidal damped function in the form
\begin{equation}\label{eq:Brillouin2}
\Delta\textrm{R}(t) \sim \sin (2\pi\nu t + \phi) \ \exp(-\Gamma t),
\end{equation}
$\phi$ being a {\color{black}constant} phase parameter. As indicated in Fig.~\ref{fig2}(c), the relevant Brillouin scattering frequency fit parameter $\nu$ changes significantly as a function of temperature. The experimental {\color{black}TDBS} temperature calibration has been further fitted by a smooth polynomial function in order to extract an even more reliable temperature behavior of the Brillouin scattering frequency of glycerol under our experimental conditions at 395~nm probe wavelength and a normal incidence. {\color{black}From the calibration curve displayed in Fig.~3(c), we can calculate the temperature rise caused by cumulative laser heating corresponding to the data of Fig. 2(c). The Brillouin frequencies extracted from Fig.~2(c) for 2.4~mJ.cm$^2$ and 12~mJ.cm$^2$ fluences respectively are 25.1~GHz and 22.7~GHz, we thus calculate a temperature rise of 3~K and 33~K respectively. The temperature uncertainty of our measurement is linked to the Brillouin frequency uncertainty. The Brillouin frequency uncertainty of a single TDBS scan, that takes about 5 seconds of acquisition time, at a given fixed temperature, is in the range of 0.1~GHz and the corresponding temperature accuracy is estimated in the range of 0.5~K. It means that our TDBS measurements can monitor dynamic temperature changes of 0.5~K at a sampling frequency of 0.1~Hz. In order to further improve the temperature accuracy of our TDBS measurements, it is required to average several TDBS scans for a better signal to noise ratio and a more accurate frequency determination.}


\section{Summary}

{\color{black}Several liquid samples, OMCTS and glycerol, have been analyzed to experimentally investigate the influence of laser-induced cumulative thermal heating effects on the liquid.} Such effects can be efficiently minimized by using a good thermally conducting substrate like sapphire or silicon. An additional decrease of the influence of cumulative thermal heating effects can be achieved with multilayer sample structure where a thermal insulating SiO$_2$ layer is added in order to shield the liquid from the laser heated metallic transducer film \cite{Chaban2017}. Under certain circumstances, alternative sample structure are required in experimental situations where even slight temperature changes have to be avoided.

The extrapolation of our results to confined liquids could shed light on the thermal properties of ultrathin liquid films \cite{Christenson1982,Heuberger2001,Perkin2012}, which is an exciting experimental challenge for the understanding of nanoscale heat transport \cite{Cahill2014,Cahill2003,Volz2016}.

\section*{ACKNOWLEDGEMENTS}

The authors acknowledge financial support from CNRS (Centre National de la Recherche Scientifique) under grant Projet International de Coop\'eration Scientifique. The authors would like to thank Lionel Guilmeau for technical support as well as Mathieu Edely for Chromium deposition.

This work was partially supported by the Department of Energy under grant No.~DE-FG02-00ER15087, National Science Foundation under grants No.~CHE-0616939 and DMR-0414895, Agence Nationale de la Recherche under grant No.~ANR-12-BS09-0031-01.


\end{document}